
\rm

{\ }

\vskip2cm

\centerline{The chemical potential of the electron gas on a one dimensional}
\vskip2mm
\centerline{lattice}
\vskip2cm
\centerline{\bf Vladan \v Celebonovi\' c}
\vskip.5cm
\centerline{Institute of Physics, P.O.B. 57, 11001 Beograd, Yugoslavia}
\vskip3cm

\noindent Abstract: The chemical potential of the electron gas on a
one-dimensional lattice is determined within the discrete Hubbard model. The
result will have applications in theoretical studies of transport
properties of quasi one-dimensional organic conductors such as the
Bechgaard salts.

\vskip2cm

\vfil\eject

Introduction

\vskip.5cm

Quasi one-dimensional (Q1D) organic metals were discovered in
1980 [1], [2], [2a]. It was soon shown [3] that the electrical
conductivity of these materials could \ not \ be \ described \ by \ the
theory of electrical conductivity of normal metals.The Q1D organic
metals \ are, (since [4]), considered \ within \ the \ framework \ of \ the
Hubbard model on a one-dimensional lattice.In spite \ of \ the \ fact
that this model is one of the simplest in statistical mechanics,it
is a subject of intensive theoretical studies.The model is
solvable in one dimension,and the chemical potential is known only for
the special case of a one-dimensional half-filled band,for which
$\mu = 0$ [5].It is usually assumed that the electrons in Q1D organic
metals form a normal Fermi liquid (however, see [6] ),which means
that

$$\rm n_k \ = \ 1 \ / \ [1+exp \ \beta (\varepsilon_k \ - \ \mu)] \eqno(1)$$

\=noindent The electronic energies are given by $\rm \varepsilon_k =
-2tcos(ks)$
[5]. All the symbols have their usual meaning, t denotes the hopping and s the
lattice constant.

The purpose \ of \ the \ present \ letter \ is \ to \ determine \ the
dependence on the band-filling and temperature of the chemical potential of
the electron gas on a one-dimensional lattice.A similar problem has been
investigated
in quantum field theory,where a special prescription had to be proposed for
introducing the chemical potential in a lattice regulated field
theory (see,for example,[15]).The knowledge of the function $\rm \mu =
\mu (n, \beta, t)$ is of paramount importance for the calculation of the
electrical
conductivity and optical response functions of correlated electron systems
[13], [14].Deviating the filling from 1/2 is a theoretical procedure for
representing the effects of doping on the material. Preliminary results of
this calculation were recently published [5a].

The calculation which will be reported in this letter is a first step
towards a theoretical study of the effects of doping on the transport
properties and optical response functions of Q1D organic metals.For a recent
example of an experimental study of these effects see [7]. Apart from
the Q1D organic metals,the calculation to be reported is of interest for
general one-dimensional models of solids ( such as the Kronig-Penney
model) [8].

\vskip1cm

The calculation

\vskip.5cm

The starting point for the determination of the chemical potential is
the following relation (for example, [9]):

$$\rm n \ = \ 2 \ \int\nolimits {1/ \ [1+exp \ \beta (\varepsilon_k \ - \
\mu)]} \
dp \ / \ (2 \pi \hbar) \eqno(2)$$

\noindent where all the symbols have their usual meaning, p = $\hbar$k, and n
denotes the number of electrons per site (n = 1 corresponds to filling 1/2).
The limits of
integration are $\rm \pm \pi \hbar / s$. Using the fact that in the 1D Hubbard
model $\rm \varepsilon_k = -2 t cos (ks)$,
it follows

$$\rm dp \ = \ \hbar \ dk \ = \ (\hbar/s) \ (4 t^2 - \varepsilon^2)^{-1/2} \
d\varepsilon \eqno(3)$$

    Inserting (3) into (2) one gets, after some algebra, that

$$\rm n \ = \ 1/(\pi s) \int\limits^{2t}_{-2t} {1/ \ [1+exp \ \beta
(\varepsilon_k \ - \ \mu)]} (4 t^2 - \varepsilon^2)^{-1/2} \
d\varepsilon \eqno(4)$$

\noindent and the problem is to find a solution of this integral equation.
Instead of
using the full rigour of the theory of integral equations (such as [10]),
the problem can be solved by a suitable developement of the Fermi function.

It has recently been shown [11] that the Fermi function can be
represented in the following form:

$$\rm 1 \ / \ [1 + exp \ \beta (\varepsilon_k - \mu)] \ = \ \Theta (\mu -
\varepsilon) \ - \ \sum\limits^{\infty}_{k=0} A_{2k+1} \beta^{-2(k+1)}
\delta^{(2k+1)} (\varepsilon - \mu) \eqno(5)$$

\noindent where \ \ \ $\rm A_{2k+1} \ = \ (-1)^{k+2} [2(2^{2k+1} -
1)\pi^{2k+2} \ / \ (2k+2)$!$] B_{2k+2}$

In eq. (5) $\Theta$ denotes the step function, $\rm \delta^{(2k+1)}
(\varepsilon -
\mu)$
are the derivatives of
the $\delta$ function and $\rm B_{2k+2}$ are Bernoulli's numbers [12].
Inserting (5) into
(4) it follows that

$$\rm n \ = \ 1 \ / \ (\pi s)  \int\limits^{2t}_{-2t} [\Theta (\mu -
\varepsilon) \ - \ \sum\limits^{\infty}_{k=0} A_{2k+1} \beta^{-2(k+1)}
\delta^{(2k+1)} (\varepsilon - \mu)] (4t^2 - \varepsilon^2)^{-1/2} \ d
\varepsilon
\eqno(6)$$

This equation can be integrated by using the following relation, valid
for an arbitrary function f(x) and its n-th order derivative:

$$\rm \int\nolimits \delta^{(n)} (x - x_0) f (x) dx \ = \ (-1)^n f^{(n)} (x_0)
\eqno(7)$$

Combining (7) and (6), and limiting (5) to terms with $\rm k \le 2$ one
finally obtains

$$\rm 1 \ / \ (\pi s) [ \pi + (\pi^2 / 6 \beta^2) \partial/\partial \varepsilon
((2t)^2
- \varepsilon^2)^{-1/2} + 7 / 360 (\pi / \beta)^4$$

$$\rm \partial^3 / \partial \varepsilon^3
((2t)^2 - \varepsilon^2)^{-1/2} + 31 / (21 \times 720) (\pi / \beta)^6
\partial^5 /
\partial \varepsilon^5 ((2t)^2 - \varepsilon^2)^{-1/2}] \ = \ n \eqno(8)$$

\noindent where $\varepsilon = \mu$ because of eq.(7).

Expression (8) can be developed up to the second order in $\mu$ around the
point $\mu$ = 0, and then solved with the following result:

$$\rm \mu = (\beta t)^6 (n s - 1) \vert t \vert \ / \ [1.1029 + .1694 (\beta
t)^2 +
.0654 (\beta t)^4] \eqno(9)$$

In the case of weak hopping,this can be developed as

$$\rm \mu = (\beta t)^6(ns-1) \vert t \vert\ [.90671 - .13940 (\beta t)^2 -
.03284 (\beta t)^4] \eqno(9a)$$

$$\rm It \ is \ obvious \ that \ \lim_{n,s\to1}{\mu} = 0, \ \ which \
means \ that \ equations \ (9) \ and \ (9a)\ reduce \ to \ the \
solution \ discussed \ in \ [5].$$
A better approximation of the function $\rm \mu = \mu (\beta ,t, n)$ can be
obtained by pushing the development of eq. (8) around the point $\mu$ = 0 to
higher orders. As a test, this development was performed to third
order. However, solutions obtained in this way are too complicated for being
applicable in calculations of the conductivity of Q1D organic conductors.

\vskip1cm

Discussion

\vskip.5cm

In this letter we \ have \ determined \ the \ dependence \ of \ the \ chemical
potential of the electron gas on a \ one-dimensional \ lattice \ on \ the \
band-filling, hopping and temperature. The calculation \ was \ performed \
within \ the discrete Hubbard model. It was motivated \ by \ the \ need \ of \
introducing \ the
doping in theoretical studies of the electrical conductivity of Q1D \ organic
metals.

The developement of eq. (8) in $\mu$ around the \ point $\mu$ = 0 \ limits \
the
applicability of eq. (9) to values of n not \ too \ different \ from \ 1.
Work \ is
already in progress aiming at an expression for the chemical potential which
could be applied at arbitrary filling. Another line of progress concerns \ the
application of the results of this letter to the analysis of the \ dependence
of the conductivity of Q1D organic conductors on the doping.

\vfil\eject

\centerline{References}
\vskip1cm

\item{[1]} K.Bechgaard,C.S.Jacoben,K.Mortensen et al,Solid State Comm.,
{\bf 33}, 1119 (1980).
\vskip2mm

\item{[2]} D.Jerome,A.Mazaud,M.Ribault and K.Bechgaard,J.Physique Lett.,
{\bf 41}, L95 (1980).
\vskip2mm

\item{[2a]} J.P.Pouget,Mol.Cryst.Liq.Cryst., {\bf 230}, 101 (1993).
\vskip2mm

\item{[3]} D.Jerome and F.Creuzet,in: Novel Superconductivity (ed.by
 S.A.Wolf and V.Z.Kresin), p.103 Plenum Press,New York (1987).
\vskip2mm

\item{[4]} P.W.Anderson,Science, {\bf 235}, 1196 (1987).
\vskip2mm

\item{[5]} E.H.Lieb and F.Y.Wu, Phys.Rev.Lett., {\bf 20}, 1445 (1968).
\vskip2mm

\item{[5a]} V. Celebonovi\'c, poster AD2Po019 at the 14 EPS-CMD General
Conference, Madrid ( Spain ) March 28.- 31.1994.
\vskip2mm

\item{[6]} J. Voit,to appear in: Proc.of the NATO-ARW "The Physics and
Mathematical Physics of the Hubbard Model", San Sebastian, ( Spain ), October
3 - 8, 1993, edited by D.K. Campbell, J.M.P. Carmelo and F. Guinea, Plenum
Press, New York, (1994).
\vskip2mm

\item{[7]} R. Kato, S. Aonuma, Y. Okano et al, Techn. Rep.of ISSP (Tokyo),
{\bf A2697}, 1 (1993). (to appear in Synth. Metals).
\vskip2mm

\item{[8]} S. Fassari, J. Math. Phys., {\bf 30}, 1385 (1989).
\vskip2mm

\item{[9]} R. P. Feynman, Statistical Mechanics: A Set of Lectures, W. A.
Benjamin Inc., Reading, Mass.(1972).
\vskip2mm

\item{[10]} E. T. Whittaker and G.N. Watson, A Course in Modern Analysis,
Cambridge Univ. Press, Cambridge (1973).
\vskip2mm

\item{[11]} V. K. Lukyanov, JINR Dubna preprint P4-93-300 (1993).
\vskip2mm

\item{[12]} M. Abramowitz and I. A. Stegun, Handbook of Mathematical Functions,
Dover Publ. Inc., New York (1972).
\vskip2mm

\item{[13]} T. Dahm, L. Tewordt and S. Wermbter, Phys. Rev., {\bf B49}, 748
(1994).
\vskip2mm

\item{[14]} M. Imada, in: "Quantum Monte Carlo Methods in Condensed Matter
Physics",pp.299-316, ed.by M.Suzuki,World Scientific,Singapore,(1993).
\vskip2mm

\item{[15]} H.J.Rothe,Heidelberg preprint HD-THEP-93-12 and references given
there (1993).

\bye